\documentclass[%
 reprint,
superscriptaddress,
nofootinbib,
 amsmath,amssymb, amsfonts
 aps,
pra,
]{revtex4-1}

\pdfoutput=1

\usepackage{color}
\usepackage{graphicx}% Include figure files
\usepackage{dcolumn}% Align table columns on decimal point
\usepackage{bm}% bold math
\begin{document}

\title{Experimental Phase Estimation Enhanced By Machine Learning}

\author{Alessandro Lumino}
\thanks{These authors contributed equally}
\affiliation{Dipartimento di Fisica, Sapienza Universit\`{a} di Roma, Piazzale Aldo Moro, 5, I-00185 Roma, Italy}

\author{Emanuele Polino}
\thanks{These authors contributed equally}
\affiliation{Dipartimento di Fisica, Sapienza Universit\`{a} di Roma, Piazzale Aldo Moro, 5, I-00185 Roma, Italy}

\author{Adil S. Rab}
\affiliation{Dipartimento di Fisica, Sapienza Universit\`{a} di Roma, Piazzale Aldo Moro, 5, I-00185 Roma, Italy}

\author{Giorgio Milani}
\affiliation{Dipartimento di Fisica, Sapienza Universit\`{a} di Roma, Piazzale Aldo Moro, 5, I-00185 Roma, Italy}

\author{Nicol\`{o} Spagnolo}
\affiliation{Dipartimento di Fisica, Sapienza Universit\`{a} di Roma, Piazzale Aldo Moro, 5, I-00185 Roma, Italy}

\author{Nathan Wiebe}
\affiliation{Quantum Architectures and Computation Group, Microsoft Research, Redmond, Washington 98052, USA}

\author{Fabio Sciarrino}
\email{fabio.sciarrino@uniroma1.it}
\affiliation{Dipartimento di Fisica, Sapienza Universit\`{a} di Roma, Piazzale Aldo Moro, 5, I-00185 Roma, Italy}

\begin{abstract}
Phase estimation protocols provide a fundamental benchmark for the field of quantum metrology. The latter represents one of the most relevant applications of quantum theory, potentially enabling the capability of measuring unknown physical parameters with improved precision over classical strategies. Within this context, most theoretical and experimental studies have focused on determining the fundamental bounds and how to achieve them in the asymptotic regime where a large number of resources is employed. However, in most applications it is necessary to achieve optimal precisions by performing only a limited number of measurements. To this end, machine learning techniques can be applied as a powerful optimization tool. Here, we implement experimentally single-photon adaptive phase estimation protocols enhanced by machine learning, showing the capability of reaching optimal precision after a small number of trials. In particular, we introduce a new approach for Bayesian estimation that exhibit best performances for very low number of photons $N$. Furthermore, we study the resilience to noise of the tested methods, showing that the optimized Bayesian approach is very robust in the presence of imperfections. Application of this methodology can be envisaged in the more general multiparameter case, that represents a paradigmatic scenario for several tasks including imaging or Hamiltonian learning.
\end{abstract}

\maketitle

{\it Introduction. --} Quantum metrology is one of the most promising applications of quantum theory \cite{Giovannetti2004,Giovannetti2006,Paris2009,Giovannetti2011,Pezze2014}, where the aim is to obtain enhanced performances in the estimation of unknown physical parameters by employing quantum resources. A notable benchmark for quantum metrology is provided by phase estimation, a task where the parameter to be measured is an optical phase embedded within an interferometric setup. In this scenario, an input probe field is prepared in a suitable state and sent through the system. The value of the phase is retrieved by measuring the field after the evolution in the interferometer, and by repeating the procedure $N$ times to perform statistical analysis. While the ultimate precision achievable with classical resources is known to be bounded by the standard quantum limit (SQL), stating that the achievable error on the unknown phase $\phi$ scales as $N^{-1/2}$ (being $N$ the number of photons), the adoption of quantum inputs can in principle improve the performances up to the Heisenberg limit (HL) \cite{Giovannetti2004,Giovannetti2006}, scaling as $N^{-1}$. Several theoretical and experimental studies \cite{Mitchell2004,Higgins2007,Pezze2008,Afek2010,Krischek2011,Wolfgramm2013,Su2017,Maccone2017,Slussarenko2017} focused on devising experimental schemes able to reach quantum enhanced performances. Furthermore, recent advances in integrated photonics has opened new possibilities for the implementation and the development of phase estimation protocols \cite{Matthews2011,Silverstone2015,Kruse2015,Chaboyer2015,Ciampini2016,Vergyris2016,Dowling2016,Atzeni2017}. In parallel, a thorough investigation has been dedicated to identifying the effect of experimental noise and losses \cite{Dorner2009,Kolo2010,Knysh2011,Escher2011}. In the scenario where the parameter to be estimated is a single phase, it is always possible to identify the optimal measurements and a suitable estimator (the latter being essentially a data processing strategy) to reach the maximum performance achievable with the chosen probe state \cite{Braunstein94,Braunstein96,Paris2009}. However, those recipes guarantee the capability of reaching the optimal error only in the asymptotic regime, thus requiring to repeat the estimation process a large number of times. Conversely, in most applications it is crucial to optimally acquire information on the unknown parameter by performing only a limited number of measurements.

\begin{figure*}[ht!]
\centering
\includegraphics[width= 0.99\textwidth]{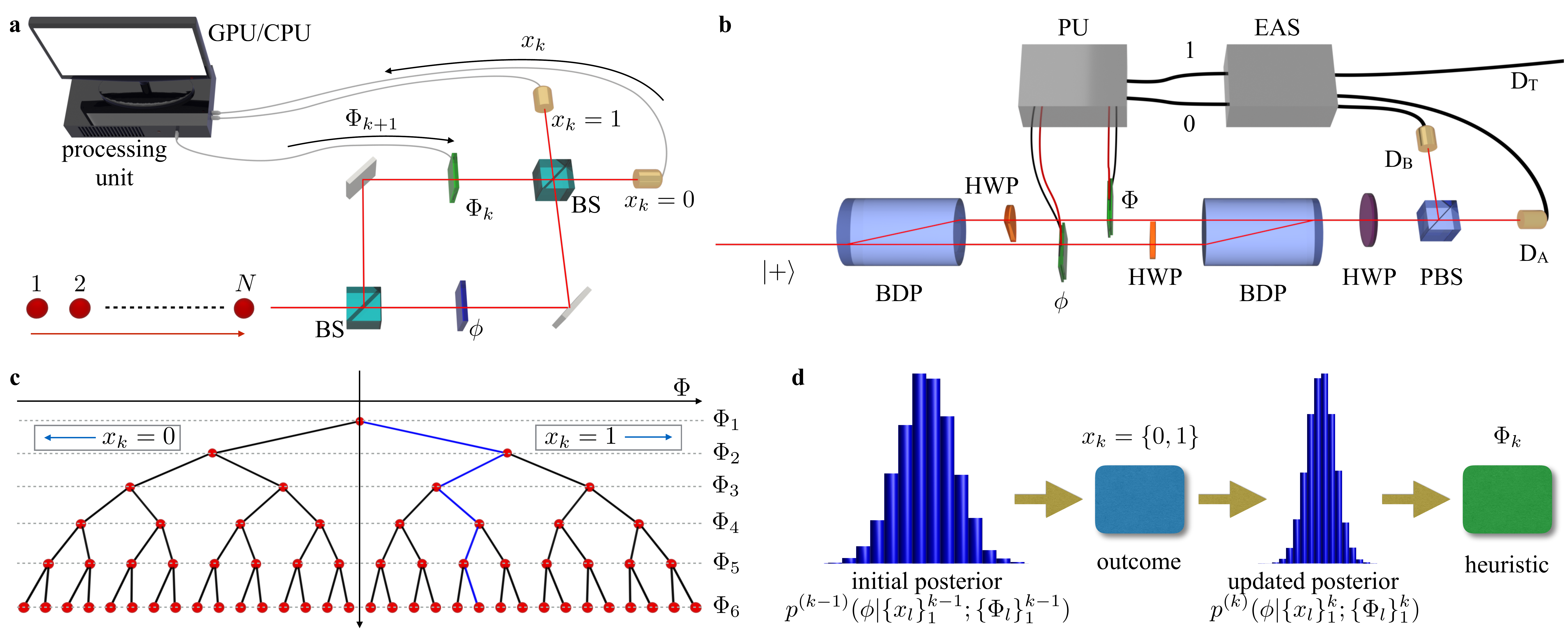} 
\caption{{\bf Adaptive-feedback scheme for phase estimation.} {\bf a}, {\it Conceptual scheme -} A two-mode Mach-Zehnder interferometer is employed as a benchmark for adaptive phase estimation protocols. At each step $k$, a single photon is injected in the interferometer and detected after the evolution. The unknown phase $\phi$ to be estimated (blue box) is kept constant during the entire experiment, while the phase $\Phi$ on the other arm (green box) is changed after each step. According to the results of the measurement $x_{k}$, the processing unit (a CPU or a GPU) properly sets a new feedback-phase $\Phi_{k+1}$ by employing a given strategy, which can be either an offline one (thus following a predetermined set of rules) or an online one (determined by the processing unit throughout the measurement process). The process is iterated until the desired number of measurements $N$ is reached. {\bf b}, {\it Experimental apparatus -} Experimental implementation of a two-mode MZI in an intrinsically stable configuration. An heralded single photon, triggered by the measurement of a photon in detector $D_{T}$ and prepared in a $\vert + \rangle = 2^{-1/2} (\vert H \rangle + \vert V \rangle)$ polarization state, is injected in the interferometer. The beam-displacing prisms (BDP) act as beam-splitters by spatially separating and recombining the optical paths depending on the photon polarization state ($\vert V \rangle$ propagates in the same mode, while $\vert H \rangle$ is displaced). Phase shifts $\phi$ and $\Phi$ are introduced in the polarization degree of freedom by means of two liquid crystal (LC) devices driven by an external processing unit (PU), one for each spatial mode (see Methods). Two half wave plates (HWP) at $45^{\circ}$ are employed within the interferometer to properly recombine the spatial modes. Finally, a HWP at $22.5^{\circ}$ and a polarizing beam-splitter (PBS) separate the two output modes which are sent to the detection stage (avalanche photo-diodes $D_{A}$ and $D_{B}$, and an electronic acquisition system, EAS). {\bf c}, {\it Decision tree for PSO-based estimation strategy -} Schematics of the binary tree at the basis of the PSO-based adaptive estimation strategy. As shown in the main text, the phase $\Phi$ is updated at each step on the basis of the last measurement result $x_{k}$ according to $\Phi_{k} = \Phi_{k-1} - (-1)^{x_{k-1}} \Delta \Phi_{k}$. Each red point in the binary tree represents a possible feedback action. The blue branch stands for one among the $2^{N}$ possible paths. {\bf d}, {\it Bayesian phase estimation protocols -} In Bayesian protocols, at each step $k$ the posterior distribution is updated according to outcome $x_{k}$, and then the new feedback phase $\Phi_{k}$ is determined from the posterior according to a chosen heuristic.}
\label{fig:figure1}
\end{figure*}

A promising approach to identify the best strategies capable of reaching optimal precision in phase estimation protocols with a few number of trials is provided by machine learning \cite{Murphy2012,Simon2013}. The latter refers to the ability of computers of learning and improving their performances without being explicitly programmed and relying mainly on experience acquired from data. Recently, several studies explored the possibility of merging the machine learning domain with the quantum world \cite{Schuld2014,Biamonte2016}. On one side, theoretical investigations have ignited the analysis on whether quantum computing can enhance machine learning protocols \cite{Lloyd2013,Rebentrost2014,Wiebe2015a,Cai2015}, for instance by providing more efficient fundamental routines. On the other side, machine learning approaches are particularly suitable to handle large amount of data and complex optimization problems, and can thus be potentially applied to improve data processing in quantum information protocols \cite{Wiebe2014a} including phase estimation \cite{Wiebe2016a,Wiebe2015b,Paesani2017}. 

In this article we report the experimental implementation of phase estimation protocols enhanced by machine learning techniques. We experimentally test for the first time an offline adaptive scheme proposed by Hentschel and Sanders in Refs. \cite{Hentschel1,HentschelProc,Hentschel2}, based on a particle-swarm algorithm, able to self-learn the optimal feedback strategy to come close to saturating fundamental limits on the scaling of the uncertainty of any unbiased estimator of the phase with the number of measurements. We then introduce a new optimized version of an adaptive  Bayesian approach, that sequentially recalculates the feedback phase according to the knowledge acquired in the previous steps and that is tailored for Gaussian prior distributions. These approaches are compared with a previously proposed adaptive Bayesian approach \cite{Wiebe2016a}, employed as a benchmark for the investigated techniques. We implement single-photon phase estimation experiments, showing the capability to reach optimal uncertainty in the parameter after a small number of trials. Furthermore, we also consider the robustness of these methods to the most relevant sources of experimental noise. These approaches can be extended in the general case where many parameters have to be estimated simultaneously, thus representing a benchmark for a significant class of learning scenarios.

{\it Adaptive phase estimation protocols. --} Adaptive protocols represent a general technique to perform phase estimation experiments starting from an unknown value of the parameter. In non-adaptive estimation protocols, the user sends $N$ times the probe state through the apparatus, and finally estimates the value of $\phi$ after collecting the full set of data. Adaptive protocols exploit additional control on the experimental system. Besides the phase $\phi$ to be estimated, the user has access to a set of physical parameters (for instance, additional phase shifts) that can be adjusted during the measurement process. After each step a single instance of the probe state is sent through the system, the set of parameters are changed by the user according to the previous knowledge acquired on the unknown phase $\phi$.

The general scheme for adaptive protocols in a two-mode Mach-Zehnder interferometer (MZI) is shown in Fig. \ref{fig:figure1}a, while the corresponding experimental apparatus is shown in Fig. \ref{fig:figure1}b. The parameter to be estimated is the phase $\phi$ corresponding to one of the two arms, while the additional parameter is provided by a feedback phase $\Phi$ inserted on the other arm of the interferometer. After each shot of the experiment, the feedback phase $\Phi$ is adjusted according to the previous knowledge acquired on $\phi$. The value of the feedback phase at each step can be determined by following either an offline or an online approach. In the first case, a processing unit determines the new value by following a pre-calculated list of rules. In the second case, the phase $\Phi$ at the subsequent stage is directly calculated step-by-step. In general, the value of the feedback phase $\Phi_{k}$ at step $k$ can be chosen starting from the results $\{x_{m}\}$ of all previous measurements ($m=1,\ldots k-1$). Hence, it is expressed by a function $\Phi_{k} = f(x_{1},\ldots x_{k-1})$ whose optimization may be in general a hard task. To this end, due to their capability to handle large amount of data and high-dimensional systems, machine learning techniques represent a promising tool to learn the optimal choice of the function $f(x_{1},\ldots x_{k-1})$. This would allow to identify the most efficient strategies able to reach the best precision on $\phi$ by using a limited number of measurements. 

{\it Particle Swarm Optimization. --} Particle Swarm Optimization (PSO) is a swarm intelligence algorithm inspired by social behavior of birds or fishes \cite{Eberhart95,Engelbrecht2006}. As birds search for food, particles search for the optimum of an objective function in $\mathbb{R}^N$. Particles are represented by a point whose coordinates defines a candidate solution to the optimization problem. The algorithm finds the optimal solution by trial and error in a fixed number of steps. Each iteration is composed by the following sequence of operations. First, every particle compares the goodness of its position, defined by a suitable measure, with respect to its previous history and to a circular neighborhood comprising the particle itself. After this comparison, the global optimum and the local optimum positions are updated accordingly. Then, each particle moves according to a given set of stochastic relations \cite{Eberhart95,Engelbrecht2006} (see Methods). After a fixed number of iterations, the algorithm returns the maximum (minimum) of the objective function. Refs. \cite{Hentschel1,Hentschel2} proposed to employ the PSO technique as a tool for a feedback-based phase estimation strategy. In their approach, the feedback phase is updated at each step from $\Phi_{k-1}$ to $\Phi_{k}$ according to the rule $\Phi_{k} = \Phi_{k-1} - (-1)^{x_{k-1}} \Delta \Phi_{k}$, where $x_{k-1}=\{0,1\}$ is the result of the measurement at step $k-1$. The final estimate for the unknown phase $\phi$ is provided by the value $\Phi_{N}$ of the adaptive phase at the end of the process. The PSO algorithm is then exploited as an offline resource to determine the set of phase shifts $\{\Delta \Phi_{k}\}$ to be applied thrughout the estimation process. A given sequence of phase shifts $\{\Delta \Phi_{k}\}$ for $k=1,\ldots N$ is named policy. The objective function optimized by the PSO algorithm measures the goodness of a policy and it is called sharpness $S(\rho) = \vert {\int_{-\pi}^{\pi} P( \theta |\rho) e^{i \theta} d \theta} \vert,$ where $P( \theta |\rho)$ represents the probability distribution of the error $\theta$ on the estimate, and $\rho$ represents a particle position, that is, a policy. This function is related to the Holevo variance $\Delta \theta_{ \rho }^{2} = S(\rho)^{-2}-1$, tailored for cyclic variables such as angles or phases. A maximum for the sharpness corresponds to a minimum in the Holevo variance for a given position of a particle. The PSO algorithm then finds the optimal policy (for a given value of $N$) by maximizing the associated sharpness (see Fig. \ref{fig:figure1}c).  Here, we apply this approach in the scenario where $N$ separable input photons are sent one by one into the input port $a$ of a MZI, thus corresponding to an input state $\vert 1 \rangle_{a}^{\otimes N}$. For each value $N$, the PSO algorithm determines its own optimal policy.

\begin{figure}[ht!]
\centering
\includegraphics[width=0.49\textwidth]{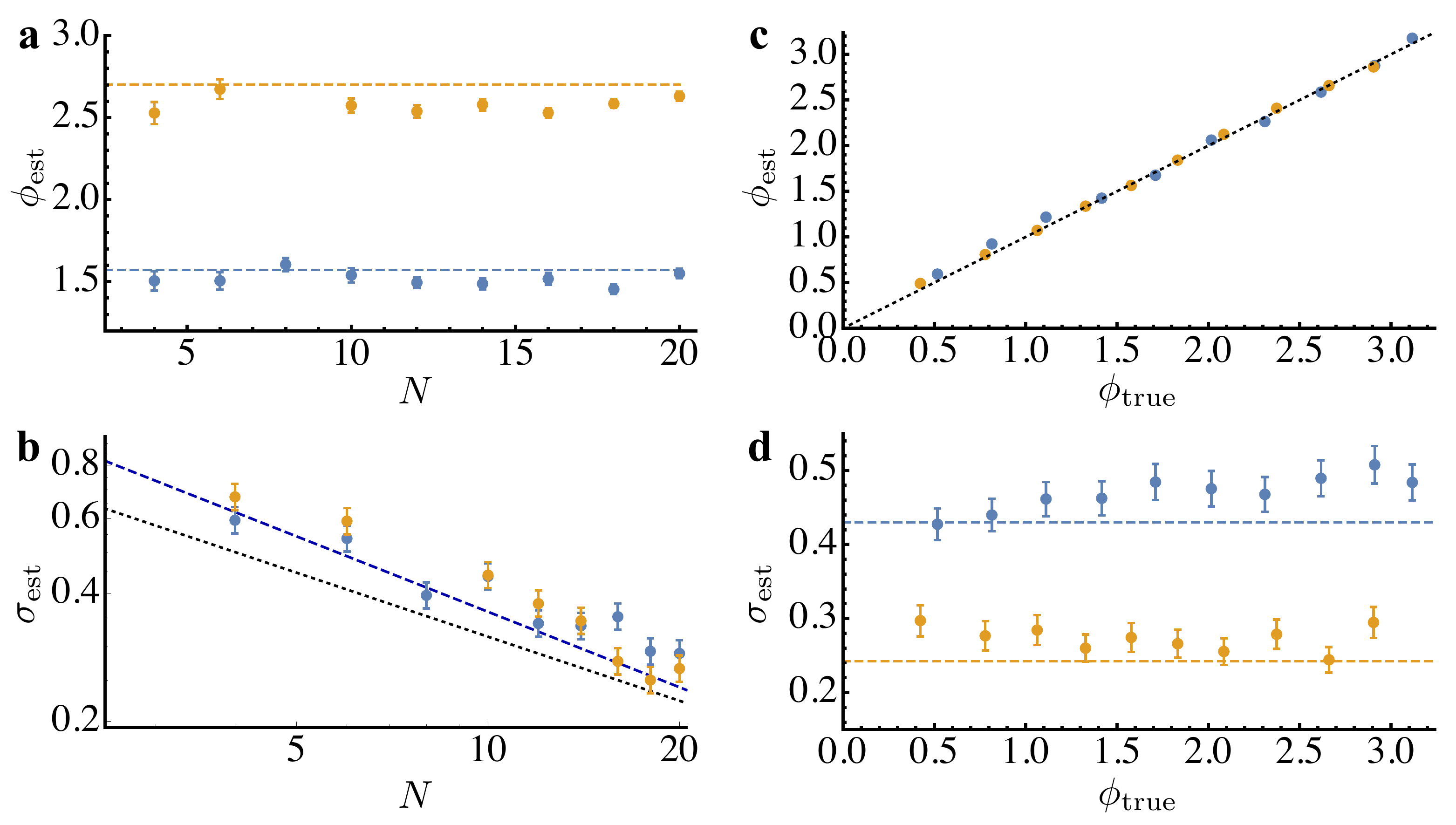}
\caption{{\bf Single-photon phase estimation measurements with the PSO approach.} Circular mean of estimated values $\phi_{\mathrm{est}}$ and square root of the Holevo variance $\sigma_{\mathrm{est}}$ obtained from $M=100$ independent experiments. {\bf a-b}, Measurements for two different phases as a function of $N$ (blue points: $\phi_{\mathrm{true}} = 1.5714$, orange points: $\phi_{\mathrm{true}} = 2.7015$). {\bf c-d}, Measurements as a function of $\phi_{\mathrm{true}}$ (blue points: $N=7$, orange points: $N=20$). 
In subfigures {\bf a,c}, bars are the errors associated to the circular mean of estimated values, evaluated as $\sigma_{\mathrm{est}} M^{-1/2}$, while dashed lines are the true value of the phases. In subfigures {\bf b,d}, bars are the errors associated to $\sigma_{\mathrm{est}}$, evaluated as $\sigma_{\mathrm{est}} [2 (M-1)]^{-1/2}$, while dashed lines are the theoretical predictions obtained from numerical simulations. In subfigure {\bf b}, black dotted line corresponds to the standard quantum limit.
}
\label{fig:figure2}
\end{figure}
\begin{figure*}[ht]
\centering
\includegraphics[width=0.99\textwidth]{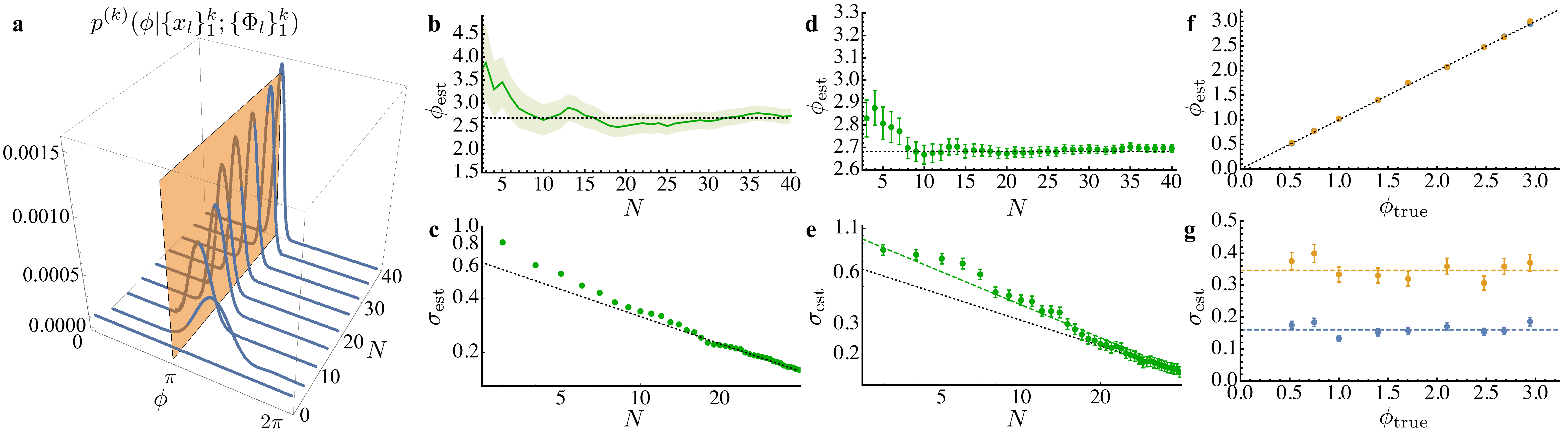}
\caption{{\bf Single-photon phase estimation experiment with the Bayesian GO approach.} {\bf a-c} Single phase estimation experiment for $\phi_{\mathrm{true}} = 2.6819$, where a sample of $N=40$ single photons are employed to obtain a final estimate. {\bf a}, Evolution of the posterior distribution $p^{(k)}(\phi \vert \{x_{l}\}_{1}^{k} ;\{\Phi_{l}\}_{1}^{k})$ during the estimation experiment. Orange plane corresponds to value $\phi_{\mathrm{true}}$. {\bf b}, Evolution of the estimated value $\phi_{\mathrm{est}}$ (green solid line) with corresponding confidence interval given by the Holevo variance of the posterior distribution (shaded region). {\bf c}, Evolution of the square root of the Holevo variance $\sigma_{\mathrm{est}}$ (green solid line) during the phase estimation experiment, compared with the standard quantum limit $N^{-1/2}$ (black dotted line). {\bf d-e}, Circular mean of estimated value $\phi_{\mathrm{est}}$ and square root of the Holevo variance $\sigma_{\mathrm{est}}$ obtained from $M=100$ independent experiments for $\phi_{\mathrm{true}}=2.6819$. In subfigure {\bf e}, green dashed line is the prediction obtained from numerical simulations, while black dotted line corresponds to the standard quantum limit. {\bf f-g}, Measurements as a function of $\phi_{\mathrm{true}}$ (blue points: $N=40$, orange points: $N=12$). In subfigures {\bf d,f} error bars are the errors associated to the circular mean of estimated values, evaluated as $\sigma_{\mathrm{est}} M^{-1/2}$, while dashed lines are the true value of the phases. In subfigures {\bf e,g} error bars are the error associated to $\sigma_{\mathrm{est}}$, evaluated as $\sigma_{\mathrm{est}} [2 (M-1)]^{-1/2}$.}
\label{fig:figure3}
\end{figure*}

{\it Bayesian phase estimation. --} Single-photon adaptive Bayesian strategies are an online approach, where the phase $\Phi$ is updated with an online method (see Fig. \ref{fig:figure1}d). Bayesian protocols start from a prior probability distribution $p^{(0)}(\phi)$, that quantifies the initial knowledge on the phase $\phi$. After each measured event $x_{k}$, the conditional probability distribution $p^{(k)}(\phi \vert \{x_{l}\}_{1}^{k} ;\{\Phi_{l}\}_{1}^{k})$ is updated according to the Bayes rule: $p^{(k)}(\phi \vert \{x_{l}\}_{1}^{k} ;\{\Phi_{l}\}_{1}^{k}) = \mathcal{N}^{-1} p(x_{k} \vert \phi; \Phi_{k}) p^{(k-1)}(\phi \vert \{x_{l}\}_{1}^{k-1} ;\{\Phi_{l}\}_{1}^{k-1})$, where $\mathcal{N}$ is a normalization constant and $p(x_{k} \vert \phi; \Phi_{k})$ is the likelihood function expressing the probability of obtaining outcome $x_{k}$ for a given value of $\phi$. For non-adaptive strategies, the phase $\Phi$ is kept constant throughout the estimation process. The conditional distribution after $N$ photons then contains all relevant statistical informations on the phase. As for the PSO approach, we assume here no a-priori knowledge on the value of the phase, thus corresponding to a prior distribution $p^{(0)}(\phi) = (2 \pi)^{-1}$.  The phase $\Phi$ is changed by the processing unit at each step $k$ according to a specific rule. The final estimated value after $N$ photons is obtained as the mean of the distribution $p^{(N)}(\phi \vert \{x_{l}\}_{1}^{N} ;\{\Phi_{l}\}_{1}^{N})$ according to $\phi_{\mathrm{est}} = \int \phi \, p^{(N)}(\phi \vert \{x_{l}\}_{1}^{N} ;\{\Phi_{l}\}_{1}^{N}) d\phi$. A benchmark choice for the feedback rule is the particle guess heuristic (PGH) approach \cite{Wiebe2016a}, where the feedback phase $\Phi$ is drawn randomly at each step from the posterior distribution $\Phi \sim p^{(k)}(\phi \vert \{x_{l}\}_{1}^{k} ;\{\Phi_{l}\}_{1}^{k})$. While this method has been proposed for Heisenberg limited metrology, it does not necessarily follow that such approaches will be appropriate in this setting where the SQL is the ultimate limit.

{\it Optimized Gaussian Phase Estimation. --}  In order to obtain an optimized Bayesian protocol in this regime, we propose below a new approach to phase estimation, named Gaussian Optimal (GO), that provides the optimal $\Phi$ under the assumptions (a) that the prior $P(\phi)$ is Gaussian with mean $\mu$ and variance $\sigma^2$ (b) that the variance of the posterior distribution obeys $\sigma \ll 0.921$ and has negligible support over the branch cut, (c) the posterior mean is used as an estimator for the phase and (d) the experimentalist wishes to minimize the expected square error in the estimate.

Under the assumption that $\sigma \ll 0.921$ and that we have negligible support over the branch cut we can write given prior mean $\mu$ and variance $\sigma^2$
\begin{equation}
P(\phi) \approx \frac{e^{-(\phi-\mu)^2/2\sigma^2}}{\sqrt{2\pi}\sigma}.
\end{equation}
Now assume that we perform an experiment and measure $x=0$ for feedback phase $\Phi$ then the posterior distribution is
\begin{equation}
P(\phi|0;\Phi) \approx \frac{\cos^2([\Phi-\phi]/2)e^{-(\phi-\mu)^2/2\sigma^2}}{\int_{-\infty}^{\infty}\cos^2([\Phi-\phi]/2)e^{-(\phi-\mu)^2/2\sigma^2}\mathrm{d}\phi}\label{eq:post0}
\end{equation}
The expected square error can then be found by computing the variance by integrating over the posterior distribution. Under these approximations this reads
\begin{equation}
\mathbb{V}(\phi|0;\Phi)=\int_{-\infty}^{\infty} \phi^2 P(\phi|0;\Phi) \mathrm{d}\phi -\left(\int_{-\infty}^{\infty} \phi P(\phi|0;\Phi) \mathrm{d}\phi \right)^2.\label{eq:variance0}
\end{equation}
Because the variance is additive, we can follow the same argument and find that the expected posterior variance that we would observe given a feedback phase $\Phi$ is chosen is
\begin{equation}
\begin{aligned}
\int_{-\infty}^\infty&\left[\cos^2([\phi-\Phi]/2))\mathbb{V}(\phi|0;\Phi)+ \right. \\
&+\left. \sin^2([\phi-\Phi]/2))\mathbb{V}(\phi|1;\Phi)\right] P(\phi)\mathrm{d}\phi,\label{eq:postvar}
\end{aligned}
\end{equation}
where $\mathbb{V}(\phi|1;\Phi)$ can be found by substituting $\cos^2(\cdot) \rightarrow \sin^2(\cdot)$ in~\eqref{eq:post0} and \eqref{eq:variance0}.

In order to find the minimum variance we then simply have to differentiate~\eqref{eq:postvar} with respect to $\Phi$ set the result equal to zero. Fortunately, analytic solutions can be found for this condition. The two corresponding to minima are:
%\begin{widetext}
\begin{equation}
\begin{aligned}
\Phi &= \mu \pm \bigg[-\pi+\\ &+\cos^{-1}\bigg(\frac{e^{\sigma^2/2}(\sigma^2 + \sqrt{(\sigma^2-2)(\sigma^2-4)}-2)}{\sigma^2-2} \bigg) \bigg)\bigg].
\end{aligned}
\end{equation}
%\end{widetext}
While the precise range of $\sigma$ for which $\Phi$ is real does not have an analytic form, we find numerically that when $\sigma\lesssim 0.921$ then these solutions are real. $\Phi=\mu$ is also a solution, but it yields a maxima not a minima of the posterior variance. 
An interesting feature of this rule is that for small $\sigma$ the feedback phase obeys $\Phi = \mu \pm (-\pi/2+\sin^{-1}(\sqrt{2}-1) +O(\sigma^2))$, which corresponds to the phase moving left or right by a constant amount based on the measurement outcome. If the particle guess heuristic (PGH) is used then we obtain random feedback phases such that $|\Phi - \mu| \sim \sigma$, which is not optimal for minimizing the quadratic loss under the assumption of a Gaussian prior with $\sigma \ll 1$ because it the optimal $\Phi$ does not converge to $\mu$ as $\sigma\rightarrow 0$. Since the branch cut can always be chosen at any point in phase, the main assumption that needs to be validated is that of the posterior being well modeled by a narrow Gaussian distribution. We provide numerical evidence for this assumption in Supplementary Note 1 and Supplementary Figure 1, while numerical simulations addressing the performance of the method for larger $N$ are shown in Supplementary Figure 2. We observe that after a few runs ($N \sim 15-20$) the estimation error reaches the standard quantum limit $N^{-1/2}$, and that such scaling is maintained for larger values of $N$. Furthermore, the algorithm allows to obtain the same performance independently from the true phase in the full $[0, 2\pi]$ interval.

{\it Experimental single-photon phase estimation. --} We have then performed adaptive single-photon phase estimation experiments by employing the different techniques discussed above. The MZI has been implemented by exploiting an intrinsically stable configuration as shown in Fig. \ref{fig:figure1}b. An external processing unit drives two liquid crystal devices that introduce the phase shifts $\phi$ (to be estimated) and $\Phi$ (the adaptive one), and is connected to an electronic acquisition system. The latter is connected to the single-photon detectors analyzing the output of the interferometer. Such system records a single event and sends the result of the measurement $(x_{m}=\{0,1\})$ to the processing unit, that changes the value of $\Phi$ according to the chosen rule. In the PSO case, phase $\Phi$ is modified according to the (precalculated) policy $\{\Delta \Phi_{k}\}$ and to the value of $x_{k-1}$ as $\Phi_{k} = \Phi_{k-1} - (-1)^{x_{k-1}} \Delta \Phi_{k}$. Conversely, in Bayesian approaches the phase $\Phi_{k}$ is calculated by the processing unit depending on the distribution $p^{(k-1)}(\phi \vert \{x_{l}\}_{1}^{k-1} ;\{\Phi_{l}\}_{1}^{k-1})$ and the chosen GO rule. This acquisition system permits to perform an actual $N$-photon experiment, where the estimate of $\phi$ is obtained by using $N$ single-photon events and no additional average is performed.

\begin{figure}[ht!]
\centering
\includegraphics[width=0.49\textwidth]{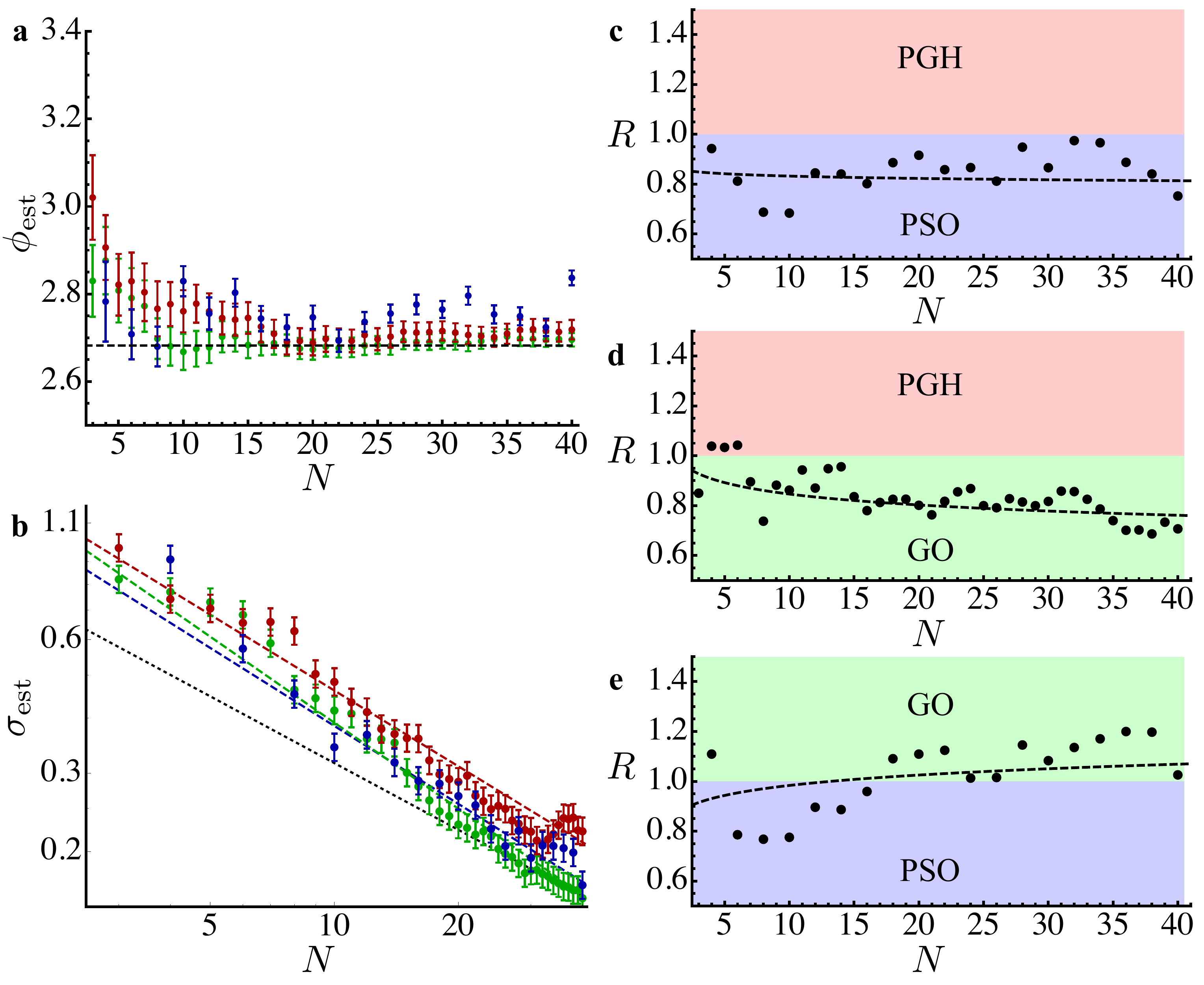}
\caption{{\bf Comparison between the different techniques}. {\bf a-b}, Circular mean of estimated value $\phi_{\mathrm{est}}$ and square root of the Holevo variance $\sigma_{\mathrm{est}}$ obtained from $M=100$ independent experiments for $\phi_{\mathrm{true}}=2.6819$ with the different techniques. Red: Bayesian PGH approach. Blue: PSO approach. Green: Bayesian GO approach. In subfigure {\bf a} error bars are the errors associated to the circular mean of estimated values, evaluated as $\sigma_{\mathrm{est}} M^{-1/2}$, while dashed lines are the true value of the phases. In subfigure {\bf b} error bars are the error associated to $\sigma_{\mathrm{est}}$, evaluated as $\sigma_{\mathrm{est}} [2 (M-1)]^{-1/2}$. {\bf c-e} Ratios between the square root of the Holevo variance for the different methods: {\bf c}, PSO and PGH, {\bf d}, GO and PGH, {\bf e}, PSO and GO. Colored region highlight which technique shows improved performances in the comparison. Dashed lines are theoretical predictions obtained from numerical simulations. Black dotted line in subfigure {\bf b} corresponds to the standard quantum limit.}
\label{fig:figure4}
\end{figure}
\begin{figure*}[ht]
\centering
\includegraphics[width=0.99\textwidth]{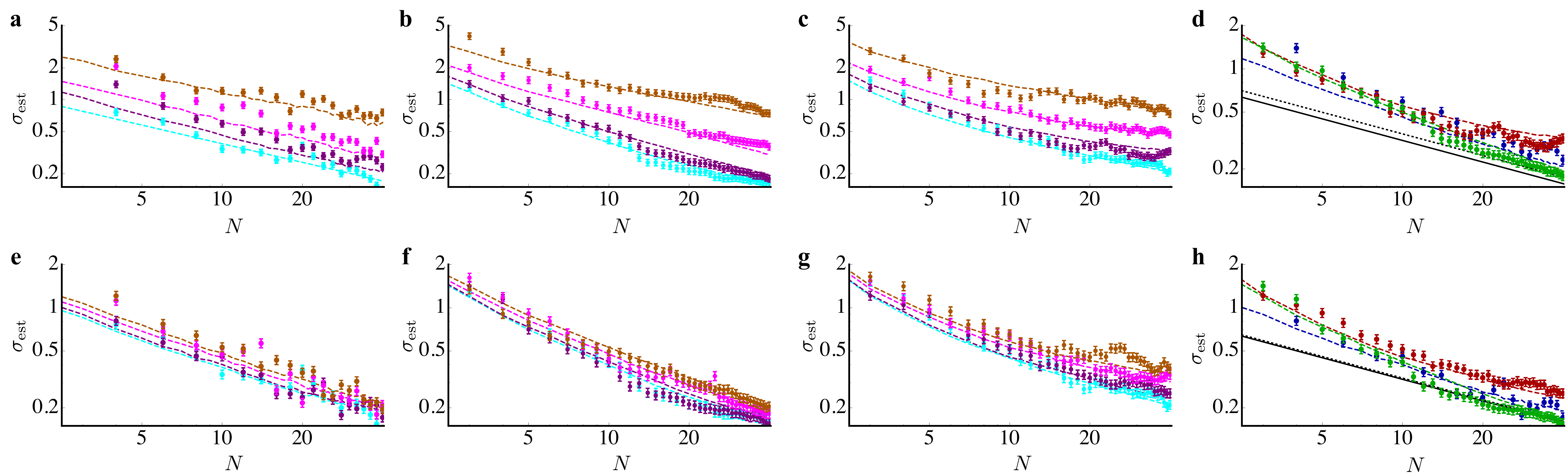}
\caption{{\bf Robustness to noise}. Plots of the square root of the Holevo variance obtained from $M=100$ independent experiments, estimated by the three approaches in presence of different values of the two verified noise models. {\bf a-d}, Results in presence of the depolarizing noise with the parameter p assuming values: 0.1 (purple points), 0.25 (magenta points) and 0.5 (orange points) compared to the noiseless case ($p=0$, cyan points). {\bf a}, PSO approach, {\bf b}, Bayesian PGH approach, {\bf c}, Bayesian GO approach and {\bf d}, comparison between the three approaches for a depolarizing parameter $p=0.1$ (blue: PSO, red: PGH, green: GO). {\bf e-f}, Results in presence of the phase noise added on the feedback phases, with the noise strength $\kappa$ assuming values: 0.2 (purple points), 0.4 (magenta points) and 0.5 (orange points) compared to the noiseless case ($\kappa=0$, cyan points). {\bf a}, PSO approach, {\bf b}, Bayesian PGH approach, {\bf c}, Bayesian GO approach and {\bf d}, comparison between the three approaches for a noise strength $\kappa=0.2$ (blue: PSO, red: PGH, green: GO). Dashed lines are theoretical predictions obtained from numerical simulations. In subfigures {\bf d} and {\bf h}, black dotted line is the optimal bound in the presence of noise, while black solid line is the standard quantum limit in the noiseless case.}
\label{fig:figure5}
\end{figure*}

The results obtained with the PSO approach are shown in Fig. \ref{fig:figure2}. The performances of the algorithm have been verified as a function of the number of photons $N$ and for different value of the phases. We performed $M=100$ independent $N$-photon experiments for each tested value of the phase ($\phi_{\mathrm{true}}$). Indeed, while our experiments yields an estimate of $\phi$ after using only $N$ photons, it is necessary to repeat the process $M$ times to evaluate the statistical error associated to the protocol (similarly to other approaches such as maximum likelihood). We observe that the experimental results are in good agreement with the predictions obtained from numerical simulations, thus showing the capability of the PSO approach to perform phase estimation experiments by using only a limited number of photons $N$.

We then performed single-photon phase estimation experiments by employing the Bayesian GO technique. The results are shown in Fig. \ref{fig:figure3}. First, we considered a single phase estimation experiment obtained by sequentially sending $N=40$ single photons through the interferometer. One of the main advantages of Bayesian techniques is the capability to provide in a single experiment an estimate of the unknown parameter and a confidence interval for the process. These informations are encoded in the posterior distributions, as shown in Figs. \ref{fig:figure3}a-c. We observe that the algorithm converges to the true value after a few runs ($N \sim 10$), and that the ultimate limit in this scenario, provided by the standard quantum limit, is reached after a limited number of measurements ($N \sim 15$). We then performed $M=100$ independent $N=40$ photons phase estimations for a fixed value of $\phi_{\mathrm{true}} = 2.6819$, showing that the protocol is stable over repeated experimental series (see Figs. \ref{fig:figure3}d-e), and the standard quantum limit can be effectively reached. Finally, the GO method is applied for different phases, showing that this approach provides similar performances independently of $\phi_{\mathrm{true}}$ (see Figs. \ref{fig:figure3}f-g).

\begin{table*}[ht!]
\centering
\begin{tabular}{||l||c|c|c||}
\hline
\hline
 &\textbf{PSO adaptive} &\textbf{Bayesian PGH} & \textbf{Bayesian GO}\\
\hline 
\emph{Type of approach} & Offline & Online & Online\\
\hline
\emph{Procedure} & Policy optimization & Posterior update & Posterior update\\
\hline
\emph{Feedback phase} & Policy and last click & Random guess & Optimized over variance\\
\hline
\emph{Computational resources in $N$} & $O(N^6)$ & $O(N)$ & $O(N)$\\
\hline
\emph{Risk Function} & Holevo variance &Mean square error & Mean square error\\
\hline
\emph{Uncertainty estimation in a single experiment} &No &Yes &Yes\\
\hline
\emph{Range of operation} &$N\lesssim 45-50$ &No limits & No limits\\
\hline
\end{tabular}
\caption{{\bf Schematic comparison between PSO heuristic and discretized Bayesian phase estimation methods.} The PSO based estimation is characterized by the offline optimization of the best policies, with risk function given by the Holevo variance. The discretized Bayesian approaches are implemented online during the experiments where, at each step, the posterior distribution is updated. The mean square error represents the corresponding risk function. These methods require different computational resource in $N$ (number of probes). The determination of the optimal policy for the PSO scales as $O(N^6)$ (when an approximation in the Holevo variance is employed). In the discretized Bayesian cases, the number of operations necessary for each step is constant, and thus a $N$-photon experiment requires $O(N)$ computational resources. The PSO estimation, unlike the Bayesian approach, does not permit to evaluate the uncertainty on the estimation in a single experiment. Finally, the PSO approach loses its effectiveness for $N\gtrsim 45-50$.}
\label{tab:comparison}
\end{table*}

Finally, in Fig. \ref{fig:figure4} we compare the performances of the PSO and the Bayesian GO approaches (see also Tab. \ref{tab:comparison}). As a benchmark, we considered the recently proposed PGH method as described above. To perform a fair comparison, Bayesian approaches have been compared to the PSO one by performing the same statistical analysis on $M=100$ independent experiments. Indeed, while in the Bayesian case it is possible to associate an estimation error to a single experiment from the distribution $p^{(N)}(\phi \vert \{x_{l}\}_{1}^{N} ;\{\Phi_{l}\}_{1}^{N})$, we performed a comparison of the two techniques by using the same analysis tools. Our test is also slightly biased in favor of PSO because PSO uses the Holevo variance as its loss function whereas the Bayesian methods are designed to minimize the quadratic loss function which need not coincide with the Holevo variance for the values of $\sigma$ tested.  We observe that both the PSO approach and the Bayesian GO one outperforms the PGH method (see Fig. \ref{fig:figure4}a-d). Finally, the Bayesian GO approach slightly outperforms the PSO one (see Fig. \ref{fig:figure4}e).

{\it Robustness to noise. --} After assessing the performances of the different algorithms, we will now examine the robustness to noise of those approaches. We considered the two most relevant sources of noise in experimental interferometric implementations, namely depolarizing and phase noise. More specifically, depolarizing noise is due to the presence of dark counts and non-unitary visibility. Conversely, phase noise corresponds to random errors on the feedback phase, which can be due to the physical device effectively introducing the phase shift or to phase fluctuations between the two interferometer arms. We have performed phase estimation experiments adding artificially those two sources of noise. Depolarizing noise can be introduced by adding a single noise parameter $p$. At each step, with probability $1-p$ the actual click $x_{k}$ of the experiment is employed, while with probability $p$ the click $x_{k}={0,1}$ is randomly drawn. Phase noise can be mimicked by adding to the feedback phase $\Phi_{k}$ a random shift $\delta \Phi$, normally distributed with zero mean and variance determined by an additional parameter $\kappa^{2}$ representing the noise strength. In both cases, the protocol is not adapted to the presence of noise, and thus neither the policies (for the PSO method) nor the likelihood function and the heuristic (for the Bayesian approaches) are modified with respect to the ideal case.

The experimental results are shown in Fig. \ref{fig:figure5}. For all techniques, we performed phase estimation experiments by considering different levels of noise for both classes. For depolarizing noise, we observe that both the PSO approach and the Bayesian techniques present a good robustness to this source of error. Furthermore, the PSO and the GO techniques still maintain a better performance than the noiseless PGH method for values of the noise parameter equal to $p \lesssim 0.1$. When phase noise is considered, we observe that these methods are very robust to this error source. Furthermore, the GO approach in the noisy case still permits to achieve lower value of the Holevo variance $\sigma_{\mathrm{est}}$ than the noiseless PGH method even when a significant amount of phase noise $\kappa \lesssim 0.5$ is introduced. For both noise models, we find that the GO approach is more robust than the PSO. This was also observed experimentally during the measurement process by noting that the PSO method has shown to be more sensitive to misalignments of the experimental apparatus. Furthermore, for moderate amount of noise the GO approach allows to reach the optimal bound with single photons when noise is taken into account (see Methods, Supplementary Note 1 and Supplementary Figure 3).

{\it Comparison with non-adaptive techniques. --} Throughout this article we have discussed the development and implementation of adaptive phase estimation techniques. Adaptive protocols have to be employed in all situations where the likelihood function present multiple values of the parameter leading to the same value of the likelihood function. Here, a two-fold periodicity is present for the phase $\phi$, that is, two different value of $\phi$ (namely, $\phi_{1}$ and $\phi_{2}$) lead to the same value of the likelihood. Conventional non-adaptive protocols are not able to distinguish between these two values ($\phi_{1}$, $\phi_{2}$), leading to an inconclusive result in the estimation process. Furthermore, adaptive strategies leads to an advantage in the estimation process also in the regime where a one-on-one correspondence between phases and likelihood is present. To this end, let us consider the interval $[0,\pi]$ for the single-phase estimation scenario discussed here. In this interval, is possible to directly apply non-adaptive estimation protocols. We compare the GO Bayesian approach with two different non-adaptive methods, by analyzing the quadratic loss $(\phi_{\mathrm{est}} - \phi_{\mathrm{true}})^{2}$ that quantifies the difference between the estimated and the true value of the phase. The first one is an inversion approach, which relies on sending $N$ single-photon probes, counting the number of times $N_{0}$ ($N_{1}$) the outcome $x=0$ ($x=1$), and estimating the phase by inverting the relation $p(0 \vert \phi; 0) \sim N_{0}/N$. The second approach is a Bayesian non-adaptive approach, which corresponds to keeping the feedback phase to $\Phi=0$ when sending $N$ independent single-photons. In Bayesian approaches, restriction to the range $[0,\pi]$ is included by truncating the prior to have support only in this interval. This corresponds to the presence of an a-priori knowledge on $\phi$. We have then performed some numerical simulation to verify the performance of the different methods. 
\begin{figure}[ht!]
\centering
\includegraphics[width=0.49\textwidth]{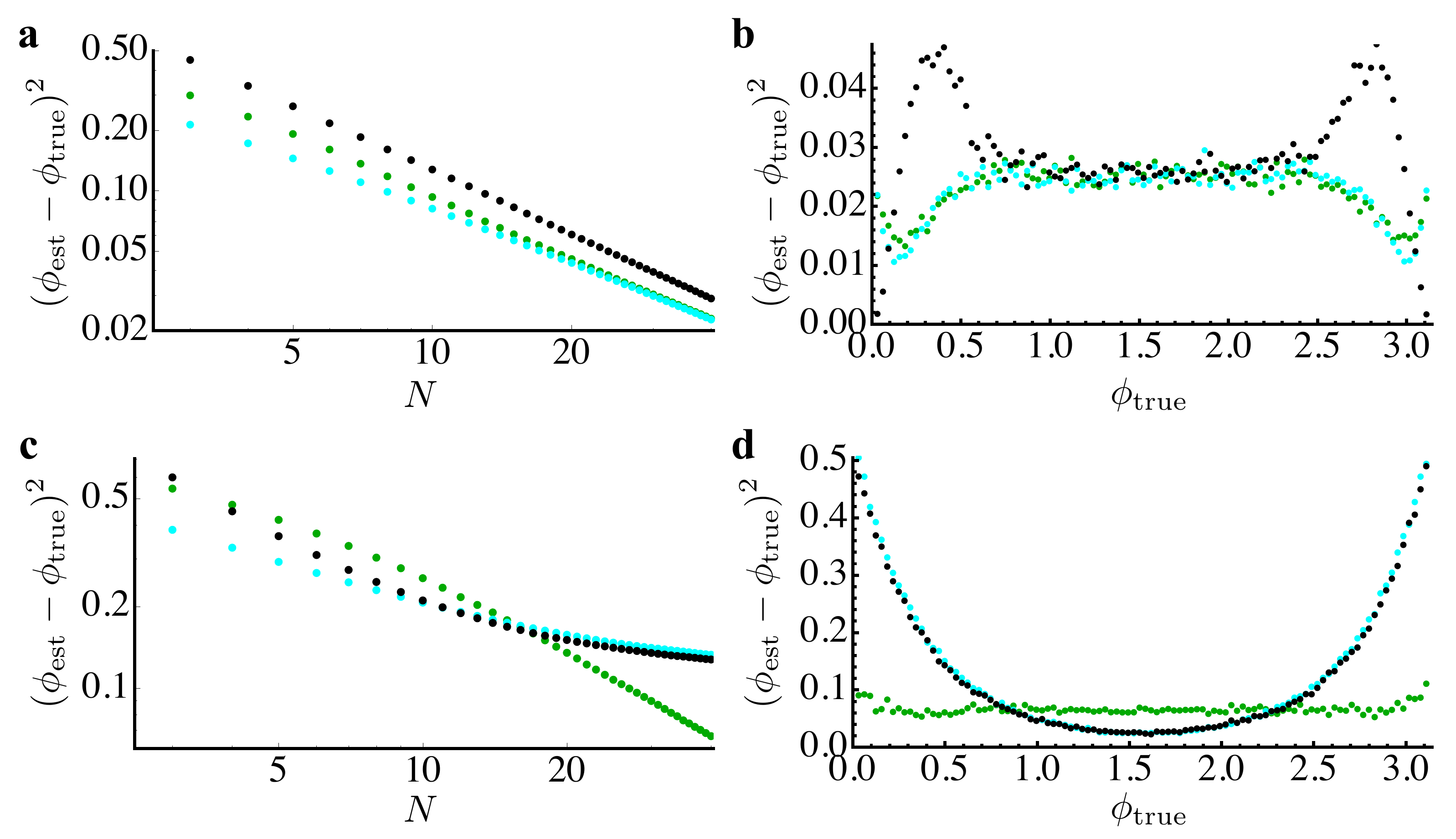}
\caption{{\bf Comparison with non-adaptive strategies.} {\bf a-b}, Quadratic loss for the estimation process in the noiseless case. {\bf c-d}, Quadratic loss in the presence of depolarizing noise with $p=0.25$. Subpanels {\bf a,c}, $(\phi_{\mathrm{est}} - \phi_{\mathrm{true}})^{2}$ as a function of $N$, averaged for 100 different phases $\phi_{\mathrm{true}}$ in the range $[0,\pi]$. Subpanels {\bf b,d}, $(\phi_{\mathrm{est}} - \phi_{\mathrm{true}})^{2}$ as a function of $\phi_{\mathrm{true}}$ for $N=40$. In all plots: green, GO method, cyan, Bayesian non-adaptive, black, inversion non-adaptive method.}
\label{fig:nonadaptive}
\end{figure}
In the noiseless case (see Fig. \ref{fig:nonadaptive}a-b), we observe that the inversion method perform worse than Bayesian approaches, also showing a phase-dependent behavior. The non-adaptive Bayesian and the GO methods present similar performances when $N \geq 10$, by considering that the GO approach is optimal for Gaussian priors (see Methods). However, the scenario changes significantly when noise is introduced in the estimation process (see Fig. \ref{fig:nonadaptive}c-d for the depolarizing case). Indeed, we observe that non-adaptive methodologies present a phase-dependent behaviour. More specifically, in the depolarizing noise case the estimation obtained with non-adaptive techniques is biased towards $\phi = \pi/2$ due to the presence of noisy random clicks. Such phase-dependent behavior is not shown by the GO approach, which also obtains better performances by averaging over $\phi_{\mathrm{true}}$. This analysis shows that adaptive methodologies perform significantly better than non-adaptive ones.

{\it Conclusions and perspectives. --} Quantum metrology protocols represent one of the most promising applications of quantum theory to improve the sensitivity in measuring unknown parameters. Phase estimation within interferometric setups can be employed as a benchmark to develop suitable techniques to be applied in the general scenario. In this article we have shown that machine learning techniques can provide a powerful tool to design optimized adaptive protocols for this purpose. We have implemented single-photon phase estimation experiments by employing different machine learning based techniques, showing the capability to reach almost optimal performances after a limited number of measurements. In particular, we have developed a new phase estimation scheme relying on Bayesian inference which is shown to saturate the standard quantum limit for very low $N$, allowing to achieve better performances that the other investigated methods. We have also shown that those techniques allow for a significant robustness to the most relevant sources of experimental noise, and that the new optimized method allows to reach optimal performances in this regime. Hence, this method can be successfully employed in a realistic scenario. 

These results highlight that machine learning methodologies can be applied to optimize the performances of quantum metrology protocols. Furthermore, those approaches can be extended to protocols exploiting quantum probes \cite{Giovannetti2006}, permitting to reach sub-standard quantum limit performances, and to the general multiparameter scenario \cite{Ciampini2016,Humphreys2013,Vidrighin2014,Datta2017,Roccia2017,Pezze2017} opening new perspectives for several applications.

{\footnotesize

\section*{Methods}

{\it Policy optimization with the PSO approach. --} The PSO approach of Refs. \cite{Hentschel1,Hentschel2} permits to learn the optimal policy for the adaptive phase estimation approach. At each step of the estimation protocol, the adaptive phase is set according to a predetermined feedback action $\{ \Delta \Phi_{k} \}$ and to the measurement outcome of the previous step $x_{k-1}$: $\Phi_{k} = \Phi_{k-1} - (-1)^{x_{k-1}} \Delta \Phi_{k}$. As described in the main text, the PSO algorithm finds the optimal policy by mapping the process to the evolution of particles in $\mathbb{R}^{N}$. At each iteration, every particle compares the goodness of its position $\rho_{i}$ respect to its previous history and to a circular neighborhood comprising the particle itself; these operations end with the updating of the general best and the local best positions. Then each particle moves according to the following relations paving the way for a new iteration: $\delta _{i} \to \beta_{1} \xi_{1} ( \hat {\rho} _{i} - \rho_{i} ) + \alpha_{2} \xi_{2} ( \hat {\Lambda} _{i} - \rho_{i} ) +\delta_{i}$ and $\rho_{i} \to \omega \delta_{i} + \rho_{i},$ where $\hat{\rho}_{i}$ is the particle best position, $ \hat {\Lambda} _{i}$ represents the local best of the $i$-th particle in its neighborhood, $\beta_{1}$ and $\alpha_{2}$ models respectively the cognitive and social behavior of the swarm; the other variables in the formulas aid PSO convergence. The operator $\to$ means that at each iteration, positions and velocities are first evaluated and then updated with the values on the right member. For further details we refer to \cite{Hentschel2} and references therein. Because of its structure, the PSO is independent of the initial state and allows intensive use of parallel GPU computing, that permits to reduce computational time with respect to a sequential evaluation. Each particle can be associated to a thread, making the entire searching process in $\mathbb{R}^N$, a simultaneous rather than a sequential one. The algorithm search for the maximum of the sharpness $S(\rho) = \vert {\int_{-\pi}^{\pi} P( \theta |\rho) e^{i \theta} d \theta} \vert$. Evaluating this function requires exponential computing time in the number of photons hence, the problem becomes rapidly intractable. To overcome this limitation, one can statistically infer the sharpness of a single particle by randomly choosing $K$ phases, estimating each one of them and then giving the sharpness estimate as $S(\rho) = \vert {\sum_{i}^{K} e^{i \varsigma_{k}}} \vert / K$, where $\varsigma_{k} = \varphi_{k} - \tilde{\varphi}_{k}$. This quantity is evaluated by the GPU associating each phase to a block and each particle to a relative thread. 

{\it Experimental details. --} Single-photon states are generated by means of a type-II spontaneous parametric down conversion process occurring in a 2 mm long beta-barium borate (BBO) crystal, pumped by a 392.5 nm, 180 fs long pulsed beam at 76 MHz repetition rate. The source generates photon pairs with orthogonal polarizations at 785 nm. Single-photon inputs are obtained by exploiting the source in a heralded configuration, thus directly detecting one of the two generated photons that acts as a trigger. Phases within the interferometer are tuned via liquid crystal devices, that change the relative phase between the horizontal and vertical polarization according to the applied electric voltage.

{\it Acquisition system. --} The electronic acquisition system is a home-built device with two output channels, that processes internally the output detected signal. When a single event is recorded, the EAS disables the acquisition process and sends the outcome of the measurement to the processing unit, consisting in a conventional desktop computer. The EAS takes as input the trigger detector $D_{T}$ and the two detectors placed at the output of the interferometer $D_{A}$ and $D_{B}$. The outcome $x_{k}=0$ ($x_{k}=1$) is obtained when a coincidence $D_{T}/D_{A}$ ($D_{T}/D_{B}$) is recorded. A LabVIEW routine then processes the input signal to determine the value of the feedback phase for the next step. In the PSO approach, the routine loads the predetermined policy and then evaluates the new value $\Phi_{k+1}$. A C-program is internally loaded to convert the phase value to the corresponding voltage for the liquid crystal device. For the Bayesian approach, an internally loaded C-program updates the conditional distribution $p^{(k)}(\phi \vert \{x_{l}\}_{1}^{k} ;\{\Phi_{l}\}_{1}^{k})$ and calculates the new value $\Phi_{k+1}$ as discussed in the main text. In both cases, the phase is then set via the LabVIEW routine within the interferometer by changing the applied voltage to the liquid crystal device. 

{\it Cram\`er-Rao bound in the presence of noise. --} The ultimate precision achievable with single-photon probes is given by the standard quantum limit, stating that the error in the estimate scales as $\delta \phi \geq N^{-1/2}$, being $N$ the number of employed probes. In the presence of noise, such limit cannot be further achieved and has to be modified to take the action of noise into account. In the general single-parameter scenario, given a specific choice of probe, evolution and measurement, the ultimate achievable precision is provided by the Cram\`er-Rao bound $\delta \phi \geq [N \mathcal{I}(\phi)]^{-1/2}$, where $\mathcal{I}(\phi)= \mathbb{E}_x [\partial \ln p(x \vert \phi)/\partial \phi]^2$ is the Fisher information associated to the output probability distribution $p(x \vert \phi;\Phi)$ at the measurement stage. 

We now evaluate the Fisher information for both noise models considered in the main text. Depolarizing noise can be modeled by a parameter $p$ corresponding to the probability of a random click. This corresponds to having output probability distributions of the form: $p_{\mathrm{dep}}(x \vert \phi; \Phi) = (1-p) p(x \vert \phi;\Phi) + p/2$, where $x=0,1$ corresponds to the two possible outcomes and $p(x \vert \phi;\Phi)$ is the output probability in the noiseless case, with $p(0 \vert \phi; \Phi) = \cos^2[(\phi - \Phi)/2]$ and $p(1 \vert \phi; \Phi) = \sin^2[(\phi - \Phi)/2]$. By directly applying the definition and by maximizing over $(\phi,\Phi)$, the Fisher information reads $\max \mathcal{I}_{\mathrm{dep}} = (1-p)^2$, and thus the minimum achievable precision is modified to: $\delta \phi_{\mathrm{dep}} \geq [N (1-p)^2]^{-1/2}$. Phase noise can be modeled by inserting a random phase shift $\delta \Phi$ in the reference arm, normally distributed with mean $\gamma=0$ and standard deviation $\kappa$ (representing the noise strength). The output probability distribution is obtained by averaging the density matrix over the random phase kick $\delta \Phi$ leading to: $p_{\mathrm{pha}}(x \vert \phi; \Phi) = e^{-\kappa^{2}/2} p(x \vert \phi;\Phi) + (1-e^{-\kappa^{2}/2})/2$. The maximum Fisher information over $(\phi,\Phi)$ reads $\max \mathcal{I}_{\mathrm{pha}} = e^{-\kappa^{2}}$, and thus the minimum achievable uncertainty on any unbiased estimator of $\phi$ (given no prior information) is modified to: $\delta \phi_{\mathrm{pha}} \geq [N e^{-\kappa^{2}}]^{-1/2}$.  

In both cases if an efficient estimator exists then the best case scenario is that such errors only impact the variance of the optimal estimator by a constant factor.  While this is indicative of the fact that such errors can be tolerated by our phase estimation protocol, they do not imply it because such a lower bound on the variance does not imply the lower bound is achievable. The fact that our GO phase estimation algorithm saturates this lower limit for moderate amount of noise, within statistical uncertainty, provides evidence in favor of the Cram\`er-Rao bound being achievable for such experiments.

\begin{acknowledgments}
This work was supported by the H2020-FETPROACT-2014 Grant QUCHIP (Quantum Simulation on a Photonic Chip; grant agreement no. 641039, http://www.quchip.eu), and by the Marie Curie Initial Training Network PICQUE (Photonic Integrated Compound Quantum Encoding, grant agreement no. 608062, funding Program: FP7-PEOPLE-2013-ITN, http://www.picque.eu).
\end{acknowledgments}

\end{document}

\title{Experimental Phase Estimation Enhanced By Machine Learning - Supplementary Information}

\author{Alessandro Lumino}
\thanks{These authors contributed equally}
\affiliation{Dipartimento di Fisica, Sapienza Universit\`{a} di Roma, Piazzale Aldo Moro, 5, I-00185 Roma, Italy}

\author{Emanuele Polino}
\thanks{These authors contributed equally}
\affiliation{Dipartimento di Fisica, Sapienza Universit\`{a} di Roma, Piazzale Aldo Moro, 5, I-00185 Roma, Italy}

\author{Adil S. Rab}
\affiliation{Dipartimento di Fisica, Sapienza Universit\`{a} di Roma, Piazzale Aldo Moro, 5, I-00185 Roma, Italy}

\author{Giorgio Milani}
\affiliation{Dipartimento di Fisica, Sapienza Universit\`{a} di Roma, Piazzale Aldo Moro, 5, I-00185 Roma, Italy}

\author{Nicol\`{o} Spagnolo}
\affiliation{Dipartimento di Fisica, Sapienza Universit\`{a} di Roma, Piazzale Aldo Moro, 5, I-00185 Roma, Italy}

\author{Nathan Wiebe}
\affiliation{Quantum Architectures and Computation Group, Microsoft Research, Redmond, Washington 98052, USA}

\author{Fabio Sciarrino}
\email{fabio.sciarrino@uniroma1.it}
\affiliation{Dipartimento di Fisica, Sapienza Universit\`{a} di Roma, Piazzale Aldo Moro, 5, I-00185 Roma, Italy}

\maketitle

\section*{Supplementary Note 1: Numerical simulations with the Gaussian optimal (GO) approach}

We have performed numerical simulations to characterize the performances of the Gaussian Optimal (GO) approach for larger values single-photon probes $N$. As discussed in the main text, this approach finds the optimal rule for updating $\Phi$ in a online Bayesian adaptive protocol to minimize the square error in the estimate.
This approach is proven to be optimal provided that (a) that the prior is Gaussian with mean $\mu$ and variance $\sigma^2$ (b) that the variance of the posterior distribution obeys $\sigma \ll 0.921$ and has negligible support over the branch cut, and (c) the posterior mean is used as an estimator for the phase. As discussed in the main text, assumption (a) is the relevant one to be validated. In Supplementary Figure 1 we provide some numerical evidence of such assumption, by showing the evolution of the posterior distribution as a function of the number of measurements. 

We have first performed some numerical simulations in the noiseless case, where the likelihood functions $p(0 \vert \phi; \Phi) = \cos^2[(\phi - \Phi)/2]$ and $p(1 \vert \phi; \Phi) = \sin^2[(\phi - \Phi)/2]$ correctly model the experimental apparatus. We observe that, after a few measurements $N \sim 20$, both quadratic loss $(\phi_{\mathrm{est}} - \phi_{\mathrm{true}})^{2}$ (Supplementary Figure 2a) and square root of Holevo variance $\sigma_{\mathrm{est}}$ (Supplementary Figure 2b) approach the standard quantum limit. Such scaling is maintained for significantly larger values of $N$, where assumption (a) is fulfilled. Furthermore, the performances of the algorithm are independent from the true phase $\phi_{\mathrm{true}}$ (Supplementary Figures 2c-d), thus showing that the applied algorithm reaches optimal and unbiased performances in the full $[0,2\pi]$ interval. 

Then, we have performed additional numerical simulations in the presence of noise, by addressing the scenarios of depolarizing and phase noise discussed in the main text. More specifically, we analyzed whether it is possible to reach the Cramer-Rao bound in this regime with the GO approach (see Methods). We observe that for moderate amount of noise ($p = 0.1$ and $\kappa = 0.2$), both quadratic loss and square root of Holevo variance approaches the Cramer-Rao bound, while this does not hold for larger noise values ($p = 0.25$ and $\kappa = 1$).  

%% Supplementary Figure 1
\begin{suppFig}[ht!]
  \centering
\includegraphics[width=\textwidth]{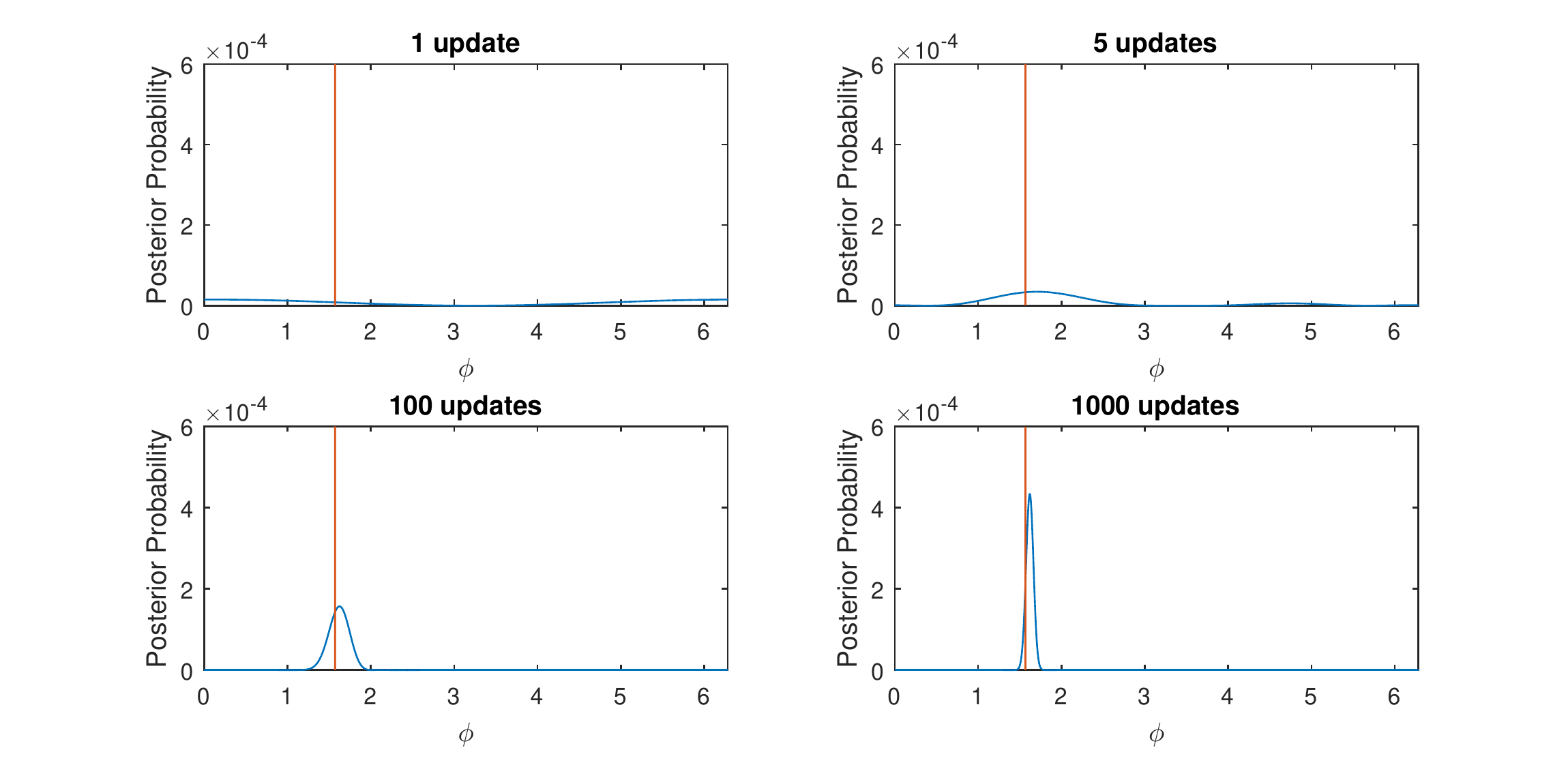}
\caption{Plot of posterior probability distribution using a $2^{17}$ point discretization of the prior probability distribution as a function of the number of measurements using the real component of the optimal $\Phi$ for a narrow Gaussian distribution under the assumption of a uniform initial prior. The red line denotes the true phase which is $\phi=\pi/2$.}
\label{fig:assumption}
\end{suppFig}

%% Supplementary Figure 2
\begin{suppFig}[ht!]
  \centering
\includegraphics[width=0.8\textwidth]{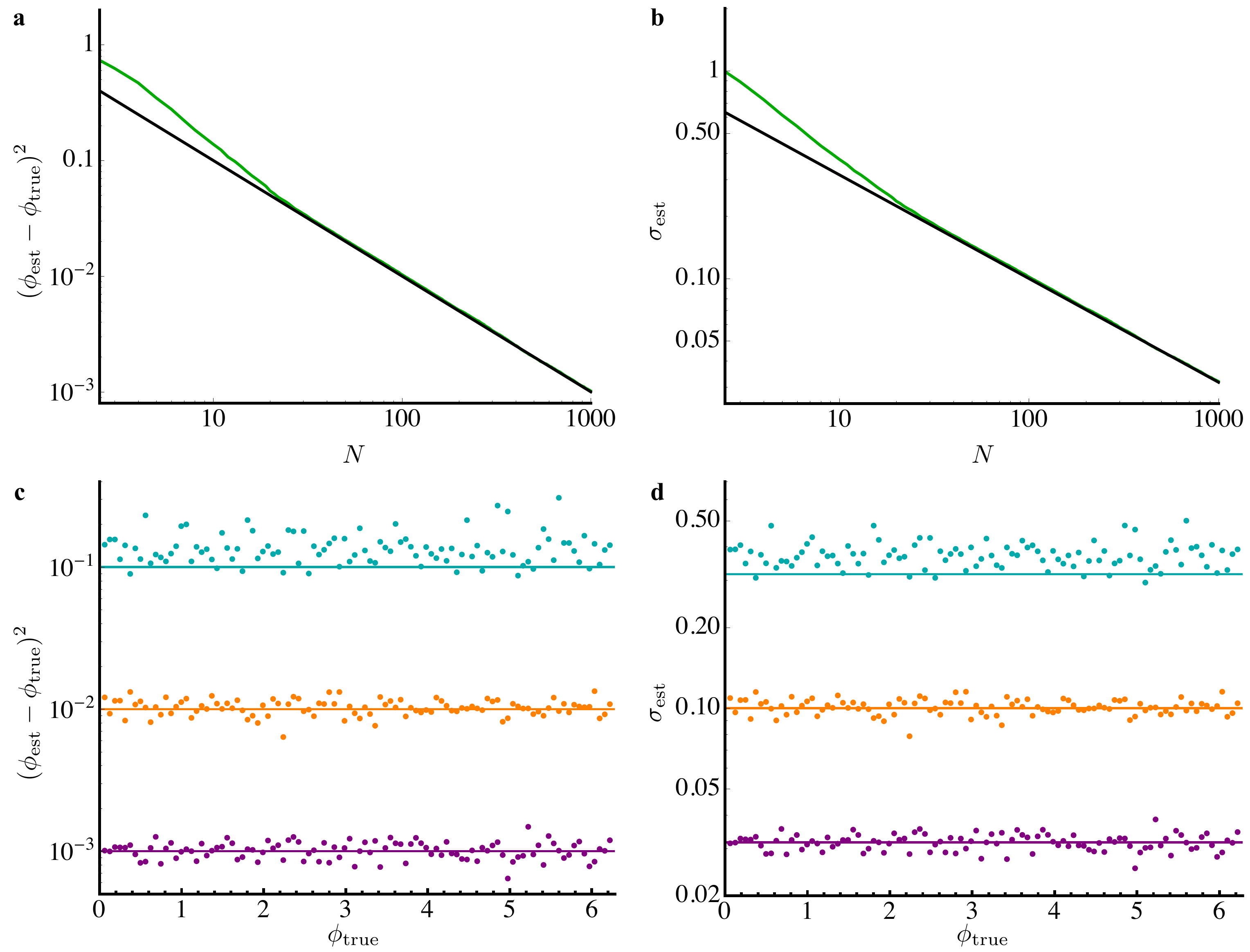}
\caption{Numerical simulations with the GO approach in the noiseless regime. $M_{\mathrm{est}}=100$ independent simulated estimation processes were performed for $M_{\mathrm{phases}}=100$ phases distributed in the full $[0,2\pi]$ interval. {\bf a}, Average quadratic loss $(\phi_{\mathrm{est}} - \phi_{\mathrm{true}})^{2}$ and {\bf b}, square root of Holevo variance $\sigma_{\mathrm{est}}$ as a function of $N$. In {\bf a,b}, green lines correspond to numerically simulated data, while black lines are the standard quantum limit $N^{-1/2}$. {\bf c}, Quadratic loss and {\bf b} Holevo variance as a function of the true phase $\phi_{\mathrm{true}}$, for different values of $N$. Cyan: $N=10$. Orange: $N=100$. Purple: $N=1000$. Solid lines correspond to the standard quantum limit for each value of $N$.}
\label{fig:noiseless}
\end{suppFig}

%% Supplementary Figure 3
\begin{suppFig}[ht!]
  \centering
\includegraphics[width=0.8\textwidth]{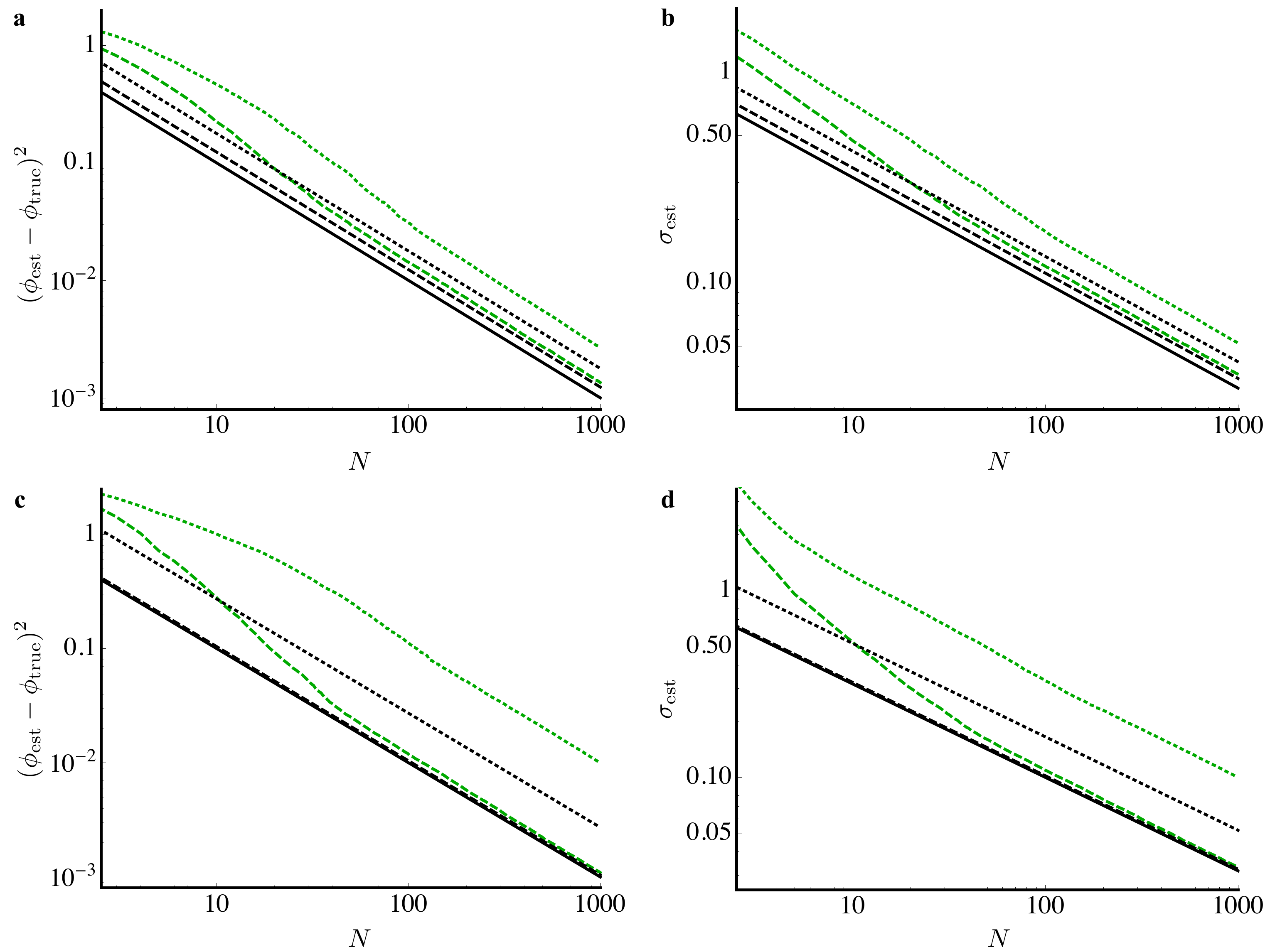}
\caption{Numerical simulations with the GO approach in the noisy regime. $M_{\mathrm{est}}=100$ independent simulated estimation processes were performed for $M_{\mathrm{phases}}=100$ phases distributed in the full $[0,2\pi]$ interval. {\bf a}, Average quadratic loss $(\phi_{\mathrm{est}} - \phi_{\mathrm{true}})^{2}$ and {\bf b}, square root of Holevo variance $\sigma_{\mathrm{est}}$ as a function of $N$ with depolarizing noise. Black solid line: standard quantum limit $N^{-1/2}$. Dashed lines: depolarizing noise with $p=0.1$ (black: Cramer-Rao bound, green: numerical simulations with GO method). Dotted lines: depolarizing noise with $p=0.25$ (black: Cramer-Rao bound, green: numerical simulations with GO method). {\bf c}, Average quadratic loss $(\phi_{\mathrm{est}} - \phi_{\mathrm{true}})^{2}$ and {\bf d}, square root of Holevo variance $\sigma_{\mathrm{est}}$ as a function of $N$ with phase noise. Black solid line: standard quantum limit $N^{-1/2}$. Dashed lines: phase noise with $\kappa=0.2$ (black: Cramer-Rao bound, green: numerical simulations with GO method). Dotted lines: phase noise with $\kappa=1$ (black: Cramer-Rao bound, green: numerical simulations with GO method).}
\label{fig:noise}
\end{suppFig}